\documentclass[twocolumn,superscriptaddress,prb,aps,floats,floatfix,showpacs,ansmath,amssymb,citeautoscript]{revtex4}
\usepackage{graphicx}
\usepackage{bm}
\usepackage{times}

\begin{document}
%
% Definitions -----------------------------------------------
%
\newcommand{\be}{\begin{equation}}
\newcommand{\ee}{\end{equation}}
\newcommand{\bea}{\begin{eqnarray}}
\newcommand{\eea}{\end{eqnarray}}
\newcommand{\beann}{\begin{eqnarray*}}
\newcommand{\eeann}{\end{eqnarray*}}
\newcommand{\bma}{\begin{array}{cc}}
\newcommand{\ema}{\end{array}}
\newcommand{\fr}{\frac}
\newcommand{\ra}{\rangle}
\newcommand{\la}{\langle}
\newcommand{\li}{\left}
\newcommand{\re}{\right}
\newcommand{\ri}{\right}
\newcommand{\uarr}{\uparrow}
\newcommand{\darr}{\downarrow}
\newcommand{\df}{\stackrel{\rm def}{=}}
\newcommand{\nn}{\nonumber}
\newcommand{\dpl}{\displaystyle}
\newcommand{\p}{\partial}

\newcommand{\alp}{\alpha}
\newcommand{\sig}{\sigma}
\newcommand{\eps}{\epsilon}
\newcommand{\xsi}{\xi}
\newcommand{\lam}{\lambda}
\newcommand{\ny}{\nu}

% paper specific:
\newcommand{\HamO}{\hat{H}_0}
\newcommand{\Ham}{\hat{H}}
\newcommand{\HamV}{\hat{V}_c}
\newcommand{\seteps}{ \{ \eps \} }
\newcommand{\setlam}{ \{ \lam \} }
\newcommand{\ef}{E_F}
\newcommand{\Deltaml}{d}%{\Delta_E}
\newcommand{\Deltaov}{\Delta}%{\Lambda}
\newcommand{\Deltamubargamma}{ \Deltaov_{\bar{\mu}\gamma} }
\newcommand{\Deltaibarj}{ \Deltaov_{\bar{i} j} }
\newcommand{\vc}{v_c}
\newcommand{\VKOSTYA}{V_K}
\newcommand{\delEF}{\delta_F}
\newcommand{\brc}{{\bf r}_c}
\newcommand{\br}{{\bf r}}
\newcommand{\omth}{\omega_{\rm th}}
\newcommand{\mata}{ {\bf a}}

%
% Title --------------------------------------------------------
%
\title{
Fermi Edge Singularities in the Mesoscopic Regime: \\
I. Anderson Orthogonality Catastrophe }

\author{Martina Hentschel}
\affiliation{
Duke University, Department of Physics, Box 90305, Durham, NC 27708-0305}
\affiliation{
Institut f\"ur Theoretische Physik, Universit\"at Regensburg, 93040
Regensburg, Germany}

\author{Denis Ullmo}
\thanks{Permanent address: Laboratoire de Physique Th\'eorique et
Mod\`eles Statistiques (LPTMS), 91405 Orsay Cedex, France}

\author{Harold U. Baranger}
\affiliation{
Duke University, Department of Physics, Box 90305, Durham, NC 27708-0305}

\date{\today}

%
%%%%%%%%%%%%%%%%%%%%%%%%%%%%%%%%%%%%%%%%%%%%%%%%%%%%%%%%%%%%%%%%%%%%
%% A B S T R A C T
%%%%%%%%%%%%%%%%%%%%%%%%%%%%%%%%%%%%%%%%%%%%%%%%%%%%%%%%%%%%%%%%%%%%
%
\begin{abstract}
For generic mesoscopic systems like quantum dots or
nanoparticles, we study the Anderson orthogonality catastrophe (AOC)
and Fermi edge singularities in photoabsorption spectra in a
series of two papers. In the present paper we focus on
AOC for a finite number of particles in discrete energy levels
where, in contrast to the bulk situation, AOC is not complete.
Moreover, fluctuations characteristic for mesoscopic systems lead
to a broad distribution of AOC ground state overlaps. The
fluctuations originate dominantly in the levels around the Fermi
energy, and we derive an analytic expression for the probability
distribution of AOC overlaps in the limit of strong perturbations.
We address the formation of a bound state and its importance for
symmetries between the overlap distributions for
attractive and repulsive potentials.  Our results are based on a
random matrix model for the chaotic conduction electrons that are
subject to a rank one perturbation corresponding, e.g., to the
localized core hole generated in the photoabsorption process.
%that we will discuss in the second paper \ \onlinecite{inprep} of this series.
%
%\onlinecite{inprep}.
% We address the mesoscopic x-ray edge problem within a Fermi golden rule
%approach and compute the average x-ray photoabsorption cross section
%for both $K$- and $L$-edge. Whereas the $L$-edge is peaked both
%for bulk-like and finite systems, the $K$-edge will go from rounded to
%slightly peaked as the system becomes coherent. The reason lies in the
%different properties of the dipole matrix elements in both situations.
%We also discuss fluctuations of the photoabsorption cross section and suggest
%experiments where the transition from a rounded to peaked $K$-edge
%and other characteristics of the mesoscopic x-ray edge problem could be measured.
\end{abstract}
\maketitle

%%%%%%%%%%%%%%%%%%%%%%%%%%%%%%%%%%%%%%%%%%%%%%%%%%%%%%%%%%%%%%%%%%%%
%% Introduction
%%%%%%%%%%%%%%%%%%%%%%%%%%%%%%%%%%%%%%%%%%%%%%%%%%%%%%%%%%%%%%%%%%%%
%%%%%%%%%%%%%%%%%%%%%%%%%%%%%%%%%%%%%%%%%%%%%%%%%%%%%%%%%%%%%%%%%%%%
\section{Introduction}
%%%%%%%%%%%%%%%%%%%%%%%%%%%%%%%%%%%%%%%%%%%%%%%%%%%%%%%%%%%%%%%%%%%%

Among the simplest many-body problems one can consider is a Fermi sea
of electrons interacting with a static impurity. Such an impurity is,
for example, suddenly formed when a core or valence electron is
excited above the Fermi energy by an incident photon in a bulk-metal
photoabsorption or photoluminescence experiment.  Naively, one would
expect a sharp onset of the absorption cross section at threshold with
a signal proportional to the density of states in the conduction band
above the Fermi energy. In experiments, deviations from this behavior
were found in the form of peaked or rounded edges both in metals
\cite{bulk_experiments_metal} and semiconductors
\cite{bulk_experiments_semi}. These so-called Fermi edge singularities
(FES) are a clear indication of the importance of many-body processes
that have been receiving enduring interest.

For (clean) macroscopic
systems, an essentially complete understanding of the mechanisms
causing the Fermi edge singularities has been obtained by the late
eighties
\cite{nozieres,mahan:book,schotteschotte,tanabe:seriesofpapers,tanabe:RMP1990}.
In particular, these works have emphasized the role of two competing
effects. The first one is the Anderson orthogonality catastrophe
(AOC), which is a property of the overlap between many-body
wavefunctions. However, in x-ray absorption and photoabsorption
processes, besides AOC there is another many-body response of the
conduction electrons to the impurity potential
\cite{mahan:book,nozieres,tanabe:RMP1990,xrayprl}, often referred to
as Mahan's exciton, Mahan's enhancement, or the
Mahan-Nozi\`{e}res-DeDominicis (MND) contribution. It counteracts the
effect of AOC in situations where the dipole selections rules are
fulfilled, and causes the photoabsorption cross section to diverge at
the threshold.

In the present series of two papers we are interested in these
many-body effects in mesoscopic systems \cite{sohn:MesoBook1997,beenakker:RMP1997,alhassid:RMP2000},
such as quantum dots or
nanoparticles, and in particular in the changes with respect to
the macroscopic (or bulk) case due to the confined geometry.  In
this first paper, we focus on the orthogonality catastrophe
mechanism for a finite system.  We will more particularly consider
the case of a contact (rank one), but not necessarily small,
perturbation, with the localized perturbation presented by the
core hole left behind in photoabsorption just being one example.
In the second paper, that will be referred to as
\ \onlinecite{inprep} in the following, we will discuss in detail
Fermi edge singularities and mesoscopic photoabsorption spectra
that depend not only on AOC but crucially on the MND contribution.

In a bulk-metal x-ray absorption experiment, the conduction electrons
respond to the abrupt, non-adiabatic appearance of a core hole
generated upon photoexcitation of an inner shell electron into the
conduction band by adjusting their single particle energies and
wavefunctions slightly. Since the core hole represents an effectively
positive charge the scattering potential will be attractive, and the
energy levels will be lowered. Although the overlap of the respective
single particle states before and after the perturbation is very close
to one, the many-body ground state overlap $\Deltaov$ tends to zero in
the thermodynamic limit, $\Deltaov \propto M^{-\eps}$ with $M$ the
number of conduction electrons and the parameter $\eps > 0$. This
effect, first noted by Anderson in 1967 \cite{anderson:PRL1967}, is
the Anderson orthogonality catastrophe. As a result of AOC the cross
section $A(\omega)$ for the absorption of photons with energy $\omega$
will be power-law suppressed near threshold, explaining the
above-mentioned Fermi edge singularity in the form of a rounded edge.

AOC plays an important role in all processes where a finite range
scattering potential suddenly arises or changes.
%One example is in optical transitions accompanied by phonon emission or
%absorption.
It is also prominent in the M\"ossbauer effect where
AOC enables the recoilless emission or absorption of $\gamma$-rays
by the whole crystal, but not by a single
nucleus\cite{moessbauer}. Another example is Kondo physics, except
that in that case the impurity is dynamic, rather than static as
considered here.

Turning from macroscopic to mesoscopic systems opens a new venue
to the study of AOC.  To start with, the confined geometry will
naturally induce new behavior associated with the mesoscopic
regime, among which are the lack of translational or rotational
invariance, the existence of interference effects, and the related
mesoscopic fluctuations \cite{sohn:MesoBook1997,beenakker:RMP1997,alhassid:RMP2000}.
Furthermore the advances in the
fabrication of mesoscopic and nanoscopic devices provide a great
flexibility in designing systems for which one can measure and
control processes dominated by AOC. One example would correspond
to the situation where an electron tunnels through a system of two
quantum dots joined via a small constriction. In that case the
boundary conditions of the electrons on the far dot are changed
once the tunneling electron enters the first dot
\cite{glazmanetal,levitov}, and as a consequence a change of one
electron appears as a sudden perturbation to the remaining
electrons. Another example is a mesoscopic tunnel junction
containing a localized impurity level such that tunneling from the
junction to the lead leaves behind a hole bound to the impurity.
The resulting power-law singularities in the current-voltage
characteristics were studied in Ref.~\ \onlinecite{matveev:larkin}
and observed in experiments with two- and three-dimensional
electrodes\cite{haug}. Large optical singularities in
one-dimensional electron gases formed in semiconductor quantum
wires were observed in Ref.~\ \onlinecite{calleja} and theoretically
explained as Fermi-edge singularities\cite{oreg:PRB1996}.

Aspects of AOC in disordered or mesoscopic systems have been addressed
in various publications \cite{kroha:PRB1992,vallejos:PRB2002,gefen:PRBR2002,matveev:PRL1998}.
AOC in disordered simple metals and its
influence on x-ray photoemission spectra (where the excited core
electron leaves the metal and the edge behavior is determined by the
AOC response alone) was studied by Chen and Kroha\cite{kroha:PRB1992}.  In
Ref.~\ \onlinecite{vallejos:PRB2002}, Vallejos and coworkers studied AOC
in chaotic mesoscopic systems for a weak perturbation
modeled by a random matrix.  They found large fluctuations of the AOC
overlap that is non-zero in general due to the incomplete
orthogonality. Small overlaps are correlated with avoided level
crossings between the highest filled and the lowest empty level. This
motivated the analytic study of a two-level model to which we refer below.  Mesoscopic conductors with diffusive disorder were
studied with respect to AOC by Gefen {\it et
al.}\cite{gefen:PRBR2002} The dependence of the average AOC-overlap
on the number $M$ of electrons was found to be dimension dependent,
and to show a stronger dependence on $M$
%than in canonical AOC
in two dimensions than in higher dimensions. In agreement with the results
by Vallejos {\it et al.} and the results we present here, a broad
distribution of overlaps is seen, with the small-overlap tail of
the distribution being related to few-level statistics near the
Fermi-energy. Another key result was that the
disorder-averaged AOC overlap depends non-mono\-tonically on the
disorder. This becomes clear in the limit of strong
disorder where the electrons are strongly localized and cannot be
sensitive to the addition of another impurity, and was confirmed in
their numerical results.

As already mentioned, here we will more specifically consider the
case of a short range (rank one), not necessarily small,
perturbation \cite{mahan:book,nozieres,tanabe:RMP1990,matveev:PRL1998}, which
applies for the description of the perturbation created by the
core hole.  We will furthermore consider a generic chaotic
mesoscopic system, such as a quantum dot or nanoparticle, for which
a random matrix theory (RMT) model of the conduction electrons
applies.  This particular regime will show some similarities with
both the diffusive systems studied in Ref.~\ \onlinecite{gefen:PRBR2002}
and the non-rank-one perturbative regime investigated by Vallejos
et al.\cite{vallejos:PRB2002}, but will also display interesting
specific properties.

Beyond AOC, Fermi-edge singularities in the mesoscopic regime were
considered in Refs.~\ \onlinecite{xrayprl}, \ \onlinecite{levitov}, and
\ \onlinecite{matveev:larkin}.
%% Abanin and Levitov consider electrons in
%% an open quantum dot\cite{levitov} (described by a scattering matrix
%% and attached to multi-channel leads) that are tunneling-connected to a
%% smaller, few-electron quantum dot. Charge switching in the open dot at
%% tunneling events causes AOC via Fermi sea shake-up. The tunneling
%% interaction can be described by a backscattering phase\cite{levitov}.
%% Backscattering leads to an enhancement of orthogonality that has the
%% same underlying physics as the disorder-enhancement in
%% Ref.~\onlinecite{gefen:PRBR2002} mentioned above. However, there is
%% another, counteracting (MND-like) contribution to the FES exponent
%% (arising from the interaction in the final state) that makes this
%% model a full x-ray edge problem. The total FES exponent can be tuned
%% to any value by varying the dot scattering parameters.
We will consider the full mesoscopic x-ray edge problem and the
resulting FES in photoabsorption spectra of quantum dots or
nanoparticles in the second paper of this series\cite{inprep}, and
concentrate on AOC in the present, first paper.

The paper is organized as follows. In Sec.~\ref{sec_aoc} we will
introduce our model of a rank one perturbation on discrete
(bulklike or chaotic) energy levels and apply it to the study of
AOC. We investigate the properties of the ground state overlap
probability distribution $P(|\Deltaov|^2)$ and show that the phase
shift at the Fermi energy determines the overlap fluctuations. The
presence of a bound state is addressed in
Sec.~\ref{sec_boundstate}, and its role for the symmetry of
overlap distributions for positive and negative perturbations is
discussed. Knowing that the origin of the fluctuations lies in the
levels nearest to the Fermi energy allows us to derive an
analytic formula for the overlap distribution in the limit of
strong perturbations in Sec.~\ref{sec_analfluctaoc}. We end this
first paper with a summary on AOC in generic mesoscopic systems.

%%%%%%%%%%%%%%%%%%%%%%%%%%%%%%%%%%%%%%%%%%%%%%%%%%%%%%%%%%%%%%%%%%%%
%% Model and AOC
%%%%%%%%%%%%%%%%%%%%%%%%%%%%%%%%%%%%%%%%%%%%%%%%%%%%%%%%%%%%%%%%%%%%
%%%%%%%%%%%%%%%%%%%%%%%%%%%%%%%%%%%%%%%%%%%%%%%%%%%%%%%%%%%%%%%%%%%%
\section{Anderson orthogonality catastrophe for mesoscopic chaotic systems}
%\section{Ground State Overlap: Anderson Orthogonality Catastrophe in
%the Mesoscopic Case}
\label{sec_aoc}
%%%%%%%%%%%%%%%%%%%%%%%%%%%%%%%%%%%%%%%%%%%%%%%%%%%%%%%%%%%%%%%%%%%%

Anderson orthogonality catastrophe refers to the vanishing of the
overlap $\Deltaov$ between the system's many-body ground states
before and after a finite range scattering potential is suddenly
applied, and is complete only in the thermodynamic limit
\cite{anderson:PRL1967}. In this Section we will discuss the
effect of a finite number of particles in a mesoscopic system as
well as the role of mesoscopic fluctuations. As the spin
variable does not play a particular role here, we shall not take
it into account (except for the discussion of the Friedel sum
rule). We therefore describe the unperturbed system by the
Hamiltonian
\begin{equation}
\HamO = \sum_{k=0}^{N-1} \eps_k c^\dagger_k c_k
\label{eq:hamH0}
\end{equation}
where the operator $c^\dagger_k$ creates a particle in the
unperturbed eigenstate $\varphi_k(\br)$ corresponding to the
eigenenergy $\eps_k$. We want to consider a situation where the
perturbing potential is created by the hole left by a core
electron excited into the conduction band, which due to screening
by the other electron is a short range potential.  We shall
therefore use for the perturbation a rank one, contact, potential
\begin{equation} \label{eq:HamV}
\HamV = \vc {\Omega} f^\dagger ({\brc}) f({\brc}) \; ,
\end{equation}
[$f ({\brc}) = \sum_k \varphi_k({\brc}) c_k$], that only
acts at the location $\brc$ of the core hole (${\Omega}$ is the
volume in which the electrons are confined, and the parameter $\vc$
defines the strength of the potential). Introducing $\tilde
c^\dagger_\kappa$ as creator of a particle in the perturbed
orbital $\psi_{\kappa}(\br)$ (we will use Greek indices %letters
to indicate the perturbed system), we obtain the diagonal form of
the perturbed Hamiltonian as
\begin{equation}
\Ham = \HamO + \HamV = \sum_\kappa  \lambda_{\kappa}
\tilde c^\dagger_\kappa \tilde c_\kappa \; .
\label{eq:ham}
\end{equation}
We will refer to the many-body, Slater-determinant ground
states of the unperturbed and the perturbed system as $| \Phi_0
\ra$ and $| \Psi_0 \ra$, respectively. Their overlap is $\Deltaov
= \la \Psi_0 | \Phi_0 \ra$. In the ground state the $M$ conduction
electrons fill the lowest of the $N$ levels, i.e.~the levels $i =
0 \ldots M-1$ [see also Fig.~\ref{fig_symmdelf}(a)]. Whenever
practical, we will use the index $i$ ($\mu$ for perturbed
levels) for levels below $E_F$ and $j$ ($\gamma$) for levels above
$E_F$, and $k$ ($\kappa$) when reference to all levels is made.

%%%%%%%%%%%%%%%%%%%%%%%%%%%%%%%%%%%%%%%%%%%%%%%%%%%%%%%%%%%%%%%%%%%%%%
\subsection{Bulk-like systems}
%%%%%%%%%%%%%%%%%%%%%%%%%%%%%%%%%%%%%%%%%%%%%%%%%%%%%%%%%%%%%%%%%%%%%%

In the bulk the orbital momentum $l$ is a good quantum number,
and, because the core hole potential (\ref{eq:HamV}) is
rotationally invariant, only the $l=0$ sector couples to the
perturbation.  Within this sector, the unperturbed energy levels
$\{ \eps_k \}$ near the Fermi energy can be taken equidistantly
spaced and, with a proper choice of the origin of the phase of the
wavefunctions we can take $\phi_k(\brc) = {\rm const.} =
1/\sqrt{\Omega}$.

In the following, we will call ``bulklike'' a model which, as the
$l=0$ sector of a bulk system, has no fluctuations  for the
eigenlevel spacing nor for the amplitude of the wavefunction at
the impurity.  We shall, however, assume finite the number of levels
within the bandwidth.   In this subsection we shall review briefly
some properties of this bulklike system. %these bulk-like systems.

\begin{figure}
\includegraphics[width=8.5cm]{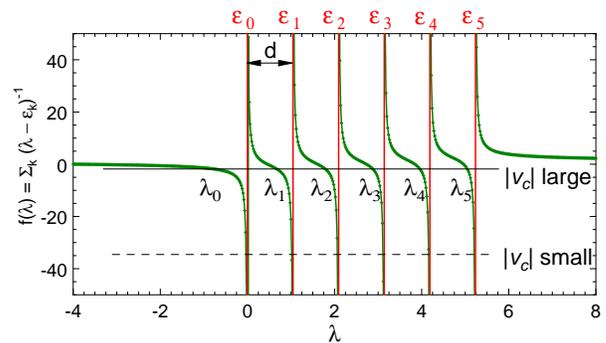}
\caption{(color online) Perturbation of a bulklike material with six
initially equidistant energy levels $\eps_k$ ($0 \le \eps_k \le 2
\pi$, mean level spacing $d$) by a rank one perturbation of strength
$\vc<0$. The perturbed eigenvalues $\lam_{\kappa}$ arise from a
graphical solution of Eq.~(\ref{eq:lambulk}) and are shifted to lower
energies (attractive interaction). This shift increases as the perturbation
becomes stronger; see the horizontal lines corresponding to
two different potential strengths. The strongest perturbation shifts
the perturbed levels right to the middle of two neighboring
unperturbed levels $\eps_k, \eps_{k+1}$ (strictly, this applies in the
center of the band -- see the discussion of boundary effects).
  \label{fig_rank1}}
\end{figure}

In the absence of mesoscopic fluctuations,  the perturbation is
\[
\HamV = \vc \sum_{kk'} c^\dagger_k c_{k'} \; .
\]
Introducing the transformation matrix ${\bf{a}} = (a_{k \kappa})$,
\begin{equation}
\psi_{\kappa} = \sum_{k=0}^{N-1} a_{k \kappa} \varphi_k \:,
\label{eq:intromata}
\end{equation}
one can check easily that the perturbed eigenstates $\{ \lambda_\kappa
\}$ and the coefficients $a_{k \kappa}$ fulfill the familiar equations
\cite{tanabe:RMP1990}
\begin{eqnarray}
0 & = & 1 - \vc \sum_{k=0}^{N-1} \frac{1}{\lam_{\kappa} - \eps_k}  \:, \label{eq:lambulk} \\
a_{k \kappa} & = & - \frac{\ny_{\kappa}}{\lam_{\kappa} - \eps_k} \:, \\
\frac{1}{|\ny_{\kappa}|^2} & = & \sum_{k=0}^{N-1} \frac{1}{\li( \lam_{\kappa} - \eps_k\re)^2 }
\label{eq:lambulk3} \:.
\end{eqnarray}
Eq.~(\ref{eq:lambulk}) allows one to find the perturbed
eigenvalues $\lam_{\kappa}$ as a function of the unperturbed
energies $\eps_k$ and the perturbation strength $\vc$. This is
illustrated in Fig.~\ref{fig_rank1} where the perturbed
eigenvalues $\lam_{\kappa}$ are found as intersection points of
the function
\[
    f(\lam) \df \sum_k \frac{1}{\lam -\eps_k}
\]
 with the horizontal line $1/\vc$. We have assumed a negative
perturbation $\vc < 0$ to account for the attractive perturbation
caused by the core hole potential.

%%%%%%%%%%%%%%%%%%%%%%%%%%%%%%%
%%%%%%%%%%%%%%%%%%%%%%%%%%%%%%%

The effect of the perturbation and the shift between perturbed and unperturbed
levels is generally characterized by the partial-wave phase shifts $\delta_l$ at
the Fermi energy for the orbital channel $l$.  The phase shifts obey the Friedel
sum rule
\begin{equation} \label{eq:Fsum}
Z=\sum_l 2 (2 l+1) \delta_l/\pi
\end{equation}
where the factor of two accounts for spin and $Z=-1$ in the case of
photoabsorption.  For a localized rank-one perturbation which, as
mentioned above, is spherically symmetric, all $\delta_l$ with $\l
\neq 0$ are zero.  In this case the Friedel sum rule
Eq.~(\ref{eq:Fsum}) implies $\delta_0 = - \pi/2$.

When $E_F$ is in the middle of the conduction band,
$\delta_0$  is related to the potential $\vc$ and the mean level
spacing $d$ by \cite{matveev:PRL1998}
\begin{equation}
\delta_0 = \arctan \frac{\pi \vc}{d}\:.
\label{eq:phaseshift_center}
\end{equation}
$\delta_0 = - \pi/2$ therefore corresponds to a strong perturbation
$|\vc/d| \!\gg\! 1$. We will therefore be mainly interested in AOC in this
limit, although we shall also consider other cases.

For the finite number of particles that we consider in the following,
boundary effects are important. First, the phase shift varies
(monotonically) with the level number, and only for levels at the band center
Eq.~(\ref{eq:phaseshift_center}) holds. For levels off-center, the number of
energy levels on the left and right is unbalanced and leads to additional
pushing from the side with the surfeit. This can already be guessed from
inspecting Fig.~\ref{fig_rank1}.

This effect can easily be taken into account by noting that,
for any perturbed level $\lambda_\kappa$ [assuming for instance $0
\leq \kappa \leq (N-1)/2$],
Eq.~(\ref{eq:lambulk}) can we rewritten as
\begin{equation} \label{eq:pushinglevels}
\sum_{k=0}^{2\kappa -1} \frac{1}{\lam_{\kappa} - \eps_k} =
\frac{1}{v_{\kappa}}
\end{equation}
where we have defined
\begin{equation} \label{eq:vkappa}
  \frac{1}{v_{\kappa}} \df  \frac{1}{\vc} - \sum_{k= 2\kappa} ^{N-1}
\frac{1}{\eps_k - \lam_{\kappa}} \:.
\end{equation}
Since $\kappa$ is now in the middle of the range $[0, 2\kappa
-1]$, we see that Eq.~(\ref{eq:phaseshift_center}) can be
generalized to give a level-dependent phase shift $\delta_\kappa =
\arctan \pi v_\kappa/d$, where evaluating the sum as an integral
yields
\begin{eqnarray}
%\fr{1}{v_i}& = & \fr{1}{\vc} + \fr{1}{\Deltaml} \ln \fr{N+0.5-i}{i+0.5}
%                   \quad \textrm {for }\;  i \in (1,N/2)    \:,\\
%\fr{1}{v_i}& = & \fr{1}{\vc} + \fr{1}{\Deltaml} \ln \fr{N-0.5-i}{i+0.5}
%                   \quad \textrm {for }\;  i \in (0,N/2-1/2)    \:,\\
\fr{1}{v_\kappa}& = & \fr{1}{\vc} + \fr{1}{\Deltaml} \ln \fr{N - \fr{1}{2} -
                   \kappa}{\kappa + \fr{1}{2}}
                   \:\: \textrm {if }\;  \kappa \in \Big( 0,\frac{N-1}{2}\Big)
\label{eq:vi} \:, \\
%\fr{1}{v_{i'}}& = & \fr{1}{\vc} - \fr{1}{\Deltaml} \ln \fr{N+0.5-i'}{i'+0.5}
%                   \quad \textrm {for }\;  i' \in (N/2,N)    \:.
%\fr{1}{v_{i'}}& = & \fr{1}{\vc} - \fr{1}{\Deltaml} \ln \fr{N+0.5-i'}{i'+0.5}
%                   \quad \textrm {for }\;  i' \in (N/2+1/2,N-1)    \:.
\fr{1}{v_{\kappa}}& = & \fr{1}{\vc} - \fr{1}{\Deltaml} \ln \fr{N - \fr{1}{2}
                   - \kappa'}{\kappa'+ \fr{1}{2}}
                   \:\: \textrm {if }\;  \kappa \in \Big(\frac{N-1}{2},N-1 \Big)    .
\nonumber %\label{eq:vigreaterNby2}
\end{eqnarray}
Note that away from the band center, $\delta_\kappa$ is no longer
confined to the interval $[-\pi/2,\pi/2]$ but can take values in
the range $[-\pi,\pi]$.

A second boundary effect becomes important for strong
perturbations $\vc/d \lesssim - 1 $ %< -0.5
that induce a large shift of the lowest perturbed energy level
towards $-\infty$ (see Fig.~\ref{fig_rank1}). We
will discuss the effect of such a bound state in detail in
Sec.~\ref{sec_boundstate}.

A  remarkable  property of rank one perturbations such as a contact potential
is that the overlap between the  many-body ground states %with $M$ particles
$|\Phi_0\ra$ and $|\Psi_0\ra$ of $\HamO$ and $\Ham$ can be expressed as
\cite{tanabe:RMP1990}
\begin{equation}
|\Deltaov|^2=
|\la \Psi_0 | \Phi_0 \ra |^2 = \prod_{i=0}^{M-1}  \prod_{j=M}^{N-1}
\fr{ (\lam_j - \eps_i)  (\eps_j - \lam_i) }{ (\lam_j - \lam_i)  (\eps_j - \eps_i)} \:.
\label{eq:overlap}
\end{equation}
In other words, all the required information is contained in the eigenvalues
$\seteps, \setlam$.  In the bulklike situation we have considered
so far, the $\setlam$ are uniquely determined from the $\seteps$,
yielding the bulklike overlap $|\Deltaov_b|^2$. Analytic
formulas for $|\Deltaov_b|^2$ were discussed in
Ref.~\ \onlinecite{tanabe:RMP1990}. For half filling, $M=N/2$, and
assuming a level-independent phase shift, the relation
\begin{equation}
|\Deltaov_b|^2 = \li( \frac{N}{2} \re)^{-\frac{\delta_0^2}{\pi^2}}
e^{-\frac{\delta_0^2}{\pi^2} {\cal F}} \:
\label{eq:overlapbulk}
\end{equation}
is easily derived. The factor ${\cal F} \approx \frac{1}{2}
\sum_{k=1}^{\infty} \frac{1}{k^2} \approx 0.822$ is a correction to
the exponent due to the discreteness of the levels.

%------- now chaotic -----------
%{\it Mesoscopic-chaotic situation.}
\subsection{Mesoscopic systems}

We now consider a generic chaotic mesoscopic system. In that case,
because of the confinement, both energies and wavefunctions will
display mesoscopic fluctuations from level to level or as an
external parameter is varied \cite{sohn:MesoBook1997,beenakker:RMP1997,alhassid:RMP2000}.
This implies that the
effective strength of the contact potential felt by electron $k$
cannot be assumed constant.  More precisely, introducing $u_k
\equiv
\sqrt{\Omega} \varphi_k(\brc) $, such that $ \langle| u_k|^2 \rangle = 1$, we
now have
\[
\HamV = \vc \sum_{kk'} u_k^* u_{k'} c^\dagger_k c_{k'} \; .
\]
Accordingly, the generalized version of
Eqs.~(\ref{eq:lambulk})-(\ref{eq:lambulk3}) for the perturbed energy
levels $\lam_{\kappa}$ reads
\begin{eqnarray}
0 & = & 1 - \vc \sum_{k=0}^{N-1} \frac{|u_k|^2}{\lam_{\kappa} - \eps_k} \:, \label{eq:lam} \\
a_{k \kappa} & = & - \frac{\ny_{\kappa} u_k}{\lam_{\kappa} - \eps_k} \:, \\
\frac{1}{|\ny_{\kappa}|^2} & = & \sum_{k=0}^{N-1} \frac{|u_k|^2}{\li(\lam_{\kappa} - \eps_k\re)^2} \:,\\
u_k^2 & = & \frac{1}{\vc}
\frac{\prod_{\kappa=0}^N \lam_{\kappa} - \eps_k}{\prod_{\kappa \neq k}^N \eps_{\kappa} - \eps_k}
\label{eq:uk} \:, \\
\tilde{u}_{\kappa} & = & \sum_{k=0}^{N-1} a_{k \kappa} u_k \equiv
-\frac{\ny_{\kappa}}{ \vc}
%%%%%%%%%%%%%%%%%%%%%%%%%%%%%%%  Denis 15/12/2004
%NB :  above, \vc used to be  2pi \vc.
%    However:
%       1) I have redefined |u_k|^2 to be N |u_k|^2 (so that
%           <|u_k|^2> = 1 instead of 1/N.
%           ---> 2pi \vc became 2pi \vc / N
%       2) 2pi here is just the bandwidth, so 2pi/N = d
%           ---> this became d \vc
%       3) The this presumably mean that this vc was the dimensionless
%         one (in units of d), when the one we have introduce in the
%         text still have energies dimensions
%           ---> this just became \vc
%%%%%%%%%%%%%%%%%%%%%%%%%%%%%%%%%%%%%%%%%%%%%%%%%%%%%%%%%%
\label{eq:uktilde} \:,
\end{eqnarray}
and the bulklike situation is recovered by setting $u_k^2 =
\langle| u_k|^2 \rangle = 1$. Note that we can choose the phase in
the wavefunctions $\varphi_k$ and $\psi_{\kappa}$ such that
their value at $\brc$ is real. It is then evident that $u_k$, $a_{k \kappa}$, and
$\tilde{u}_{\kappa}
% = \sum_k a_{k \kappa} u_k
$ are also real.

For a chaotic system, a good model for the eigenvalues and
eigenvector fluctuations is given by the classic ensembles of
Random Matrix Theory (RMT) \cite{Oriol:Houches,mehta}. Most
commonly the Gaussian orthogonal (unitary) ensembles, GOE (GUE),
are used to describe the energy eigenvalues of chaotic systems in
the presence (absence) of time reversal symmetry. Here however,
because we would like to model systems with constant mean density
of states, we will rather consider the corresponding Circular
Ensembles COE and CUE \cite{mehta} which display the same
fluctuations but have by construction a mean density of
eigenphases $\exp(i \epsilon_k)$ that is uniformly
distributed on the unit circle.  Since time reversal symmetry can
easily be broken by applying a magnetic field, we will address
both ensembles.

The fluctuations in the $\seteps$ and $\setlam$ are not
independent. Under the hypothesis that the unperturbed
eigenenergies are distributed according to the circular ensembles,
and accordingly that the eigenvectors follow a Porter-Thomas
distribution, the joint probability distribution
$P(\seteps,\setlam)$ was shown by Aleiner and Matveev
\cite{matveev:PRL1998} to be
\begin{equation}
  P(\seteps,\setlam) \propto
  \frac{\prod_{i>j} (\eps_i - \eps_j) (\lambda_i - \lambda_j)}
       {\prod_{i,j} \left|\eps_i - \lambda_j \right|^{1-\beta/2}}
       e^{ - \frac{\beta}{2} \sum_i
             (\lambda_i - \eps_i)/v_i }
  \label{eq:kostya}
\end{equation}
with the constraint $\eps_{i-1} \le \lambda_i \le \eps_i$ and $\beta = 1$ in the
COE case whereas $\beta=2$ for CUE. In the middle of the band $v_i =\vc$, away
from the band center, boundary effects have to be taken into account according
to Eq.~(\ref{eq:vi}). It is interesting that a joint probability
distribution can also be found in the more general, non-rank-one case
\cite{smolyarenko:PRL2002} that we do not consider here.

%%%%%%%%%%%%%%%%%%%%%%%%%%%%%%%
\begin{figure}
\includegraphics[width=8.5cm]{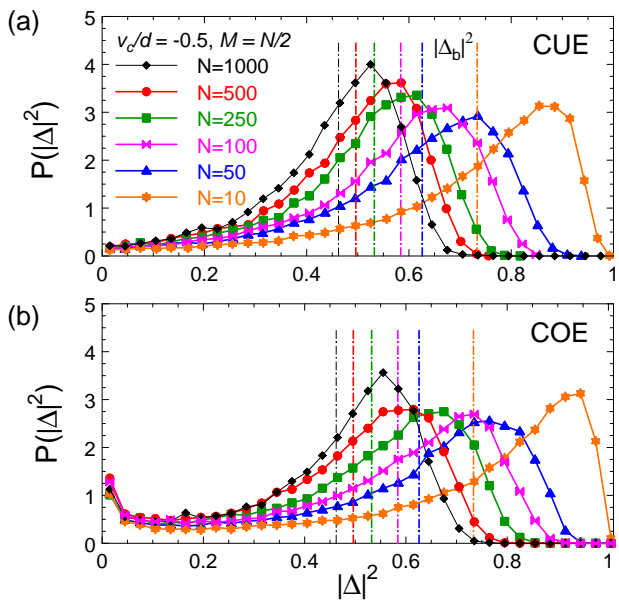}
\caption{(color online) Distribution of ground-state overlaps
in a mesoscopic system as the number of particles is increased at
half-filling, $M=N/2$, for the (a) CUE and (b) COE cases. The
perturbation strength is chosen intermediate, $\vc/d=-0.5$. The
dash-dotted lines indicate the bulklike overlaps $|\Deltaov_b|^2$.
The curves illustrate the power-law characteristic of AOC that
makes it a truly thermodynamic-limit effect: even with $M =1000$
electrons in the system, the most probable overlap $|\Deltaov|^2$
is as big as 0.46, indicating that a bulk-based discussion of AOC
in mesoscopic systems can be inaccurate. Note that for this
intermediate perturbation strength the overlap distribution
exhibits a clear maximum and a difference in CUE vs.\ COE for
small overlaps. (In order to make the statistical error
essentially negligible, 20000 realizations of the random matrix
were used in producing each curve.)
  \label{fig_aoc}}
\end{figure}
%%%%%%%%%%%%%%%%%%%%%%%%%%%%%%%

%%%%%%%%%%%%%%%%%%%%%%%%%%%%%%%
\begin{figure}
\includegraphics[width=8.5cm]{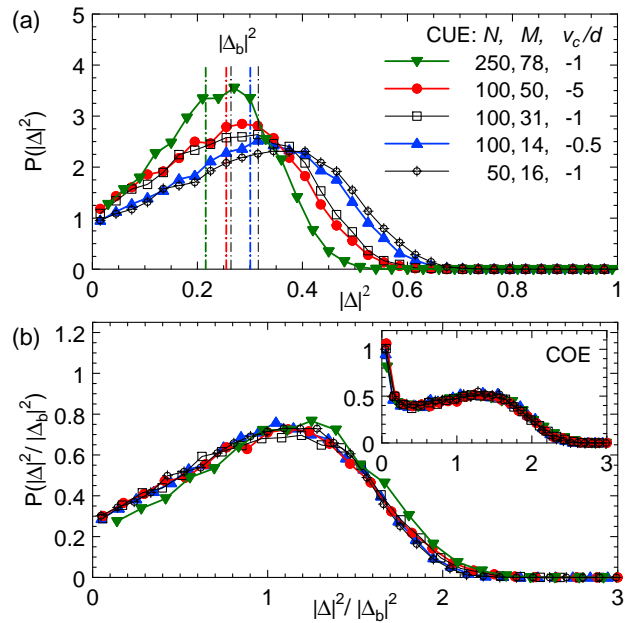}
\caption{(color online) Ground state overlap in the absence of time-reversal
symmetry (CUE case) for different parameter realizations
corresponding to a phase shift of approximately $-\pi/2$ at the Fermi energy.
(a) Before, (b) after scaling with the bulklike result
$|\Deltaov_b|^2$ (dash-dotted lines, $|\Deltaov_b|^2$ increases
with the legend entries). The scaled result for COE is given in
the inset of (b), see Ref.~\ \onlinecite{xrayprl} for the unscaled
result in that case.
  \label{fig_overlapgue}}
\end{figure}
%%%%%%%%%%%%%%%%%%%%%%%%%%%%%%%

The ground state overlap $\Delta$ is expressed entirely in terms
of the unperturbed and perturbed eigenvalues $\seteps$ and
$\setlam$ [see Eq.~(\ref{eq:overlap})].  Therefore, the
joint distribution (\ref{eq:kostya}) contains all the information
required to build the overlap distribution $P(|\Delta|^2)$.

In Section \ref{sec_analfluctaoc}, we shall see that a slightly
different route is actually more effective to obtain analytic
expressions for $P(|\Delta|^2)$.  However, the present approach is
perfectly adapted to constructing numerically this distribution for
any value of the parameters $M$, $N$, $\vc$, and $\beta$.  Indeed,
we can use Eq.~(\ref{eq:kostya}) as the basis of a simple
Metropolis algorithm to generate ensembles of $(\seteps,\setlam)$
with the proper joint probability distribution.  Applying
Eq.~(\ref{eq:overlap}), we then build the distribution of ground
state overlaps.  Examples for overlap distributions obtained in
this way are shown in Fig.~\ref{fig_aoc} for systems with and
without time-reversal symmetry and an intermediate perturbation
strength. The evolution of AOC as the number of particles is
increased is clearly visible as a shift of the probability density
to smaller values of $|\Deltaov|^2$. However, since AOC is a
power-law effect even $N=1000$ is far away from the thermodynamic
limit, cf.~Eq.(\ref{eq:overlapbulk}) that can also be used as a
rough estimate for the mean overlap in the mesoscopic case. We
point out the different behavior for very small overlaps in %between
the CUE vs.~COE case. It originates in the difference of the
Porter-Thomas distributions and will be discussed in
Sec.~\ref{sec_analfluctaoc}.

In the metallic x-ray edge problem all relevant properties are
known to depend only on the phase shift at the Fermi energy, and a
natural question is whether this remains true in the
mesoscopic case for the groundstate overlap.  In
Fig.~\ref{fig_overlapgue}(a) we therefore compare overlap
distributions for different parameter sets $\{N,M,\vc/d\}$ all of which yield
a phase shift $\delta_F \!\approx$
$\arctan (\pi v_{i=M} / d) \!\sim$ $-\pi/2$ at the Fermi energy $E_F$.
The curves differ, as do the corresponding bulklike overlaps
$|\Deltaov_b|^2$ indicated by the vertical lines. However, after
scaling by the bulklike overlaps all curves coincide,
Fig.~\ref{fig_overlapgue}(b).
Therefore, whereas the individual
$P(|\Deltaov|^2)$ depend independently on $N$, $M$, and $\vc/d$, {\it the fluctuations of the overlap depend on the
value of $\delta_F$ alone.}

\begin{figure}[t]
\includegraphics[width=8.5cm]{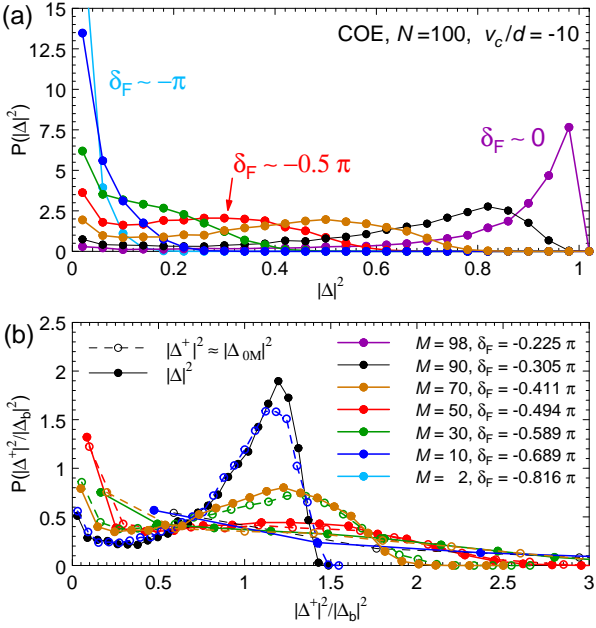}
\caption{(color online) (a) Probability distribution $P(|\Deltaov|^2)$
for the ground state overlap (COE, $N=100, \vc/d=-10$). The number
$M$ of electrons is varied, corresponding to a varying phase shift
$\delta_F$ at the Fermi energy. The curves (here not scaled by the
bulklike overlaps) illustrate the characteristics of
$P(|\Deltaov|^2)$ for a given phase shift. For comparison, recall that at constant
filling, a decreasing phase shift
corresponds to increasing perturbation strength. (b) Comparison of
the bulk-scaled probability distributions of the ground state
overlaps for a negative (solid line, $|\Deltaov|^2$) and positive
(dashed line) perturbation. The overlap $|\Deltaov^+|^2$ is, up to
the negligible influence of the run-away states, equivalent to the
overlap  $|\Deltaov_{0M}|^2$ shown here as dashed line. %and that
It will be of importance in the discussions of the x-ray
absorption part. The pairing of curves whose phase shifts moduli
add up to $\pi$ is nicely illustrated in this graph.
%The minor deviations stem mainly from the coarseness of the levels.
  \label{fig_symm_GOE}}
\end{figure}

\begin{figure}[t]
\includegraphics[width=8.5cm]{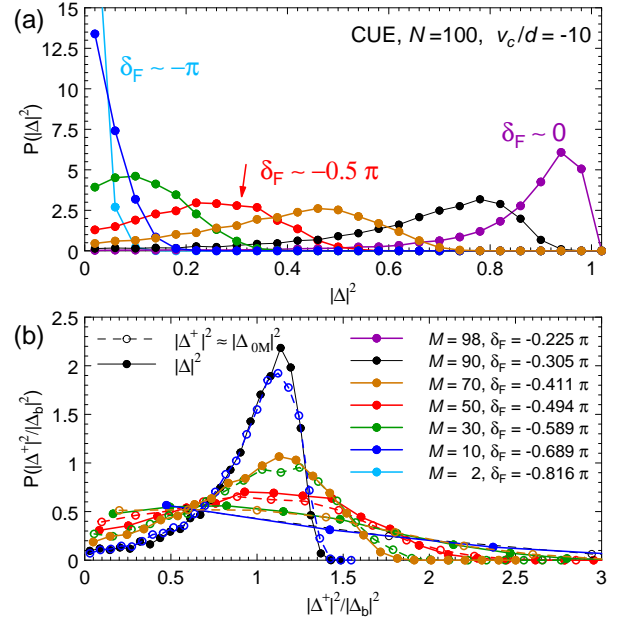}
\caption{(color online) Same as in Fig. \ref{fig_symm_GOE}, but for CUE. Note the
difference in the distributions for small overlaps $|\Deltaov|^2$.
  \label{fig_symm_GUE}}
\end{figure}

This fact allows one to provide the overlap distribution as a
universal curve for a given phase shift $\delta_F$.
%at the Fermi energy.
In Figs.~\ref{fig_symm_GOE}(a) and \ref{fig_symm_GUE}(a)
results are shown for $N=$ 100 levels subject to a strong
perturbation. The phase shift $\delta_F$ is varied by increasing
the filling of the band from 2 to 98 electrons. The distributions
are {\it not} scaled by the bulk overlap in order to allow
easier comparison of the curves when they all range from 0 to 1.

%%%%%%%%%%%%%%%%%%%%%%%%%%%%%%%
%%%%%%%%%%%%%%%%%%%%%%%%%%%%%%%%

%%%%%%%%%%%%%%%%%%%%%%%%%%%%%%%
%%%%%%%%%%%%%%%%%%%%%%%%%%%%%%%%

%%%%%%%%%%%%%%%%%%%%%%%%%%%%%%%%%%%%%%%%%%%%%%%%%%%%%%%%%%%%%%%%%%%%%%%%%%%%%
%\subsection{Applicability of the model: Kolmogorov-Smirnov test}
%\subsection{Thouless energy vs.~Kolmogorov-Smirnov test}
\subsection{Fluctuations of levels beyond the Thouless energy}
\label{sec_kstest}
%%%%%%%%%%%%%%%%%%%%%%%%%%%%%%%%%%%%%%%%%%%%%%%%%%%%%%%%%%%%%%%%%%%%%%%%%%%%%%

\begin{figure}
\includegraphics[width=8.5cm]{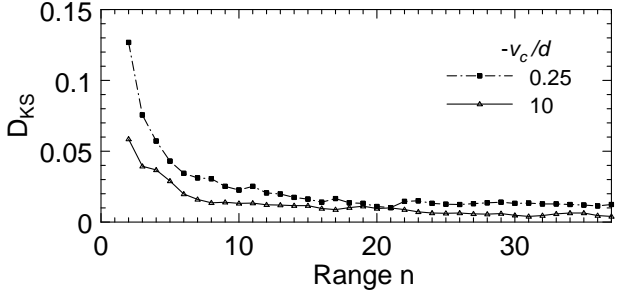}
\caption{Accuracy of the range-$n$ approximation using the Kolmogorov-Smirnov test (COE, $N=100, M=50$). The maximum
deviation of the range-$n$ cumulative distribution from the exact
cumulative distribution of the overlap is shown as a
function of $n$. The rapid approach to 0, especially in the strong perturbation case, shows that fluctuation of only those levels near $E_F$ is important.
  \label{fig_kstest}}
\end{figure}

Strictly speaking a random matrix model applies only to chaotic
energy levels on a scale of the Thouless energy $E_{\rm Th}$ \cite{berry1985,altshuler1986,beenakker:RMP1997,alhassid:RMP2000}.  In
the semiclassical limit $E_{\rm Th}$ is much larger than
the mean level spacing, but significantly smaller than the Fermi
energy $E_F$. Therefore one may question the model we use where
the full bandwidth is described by RMT.  It turns out, however, that
the overlap distributions depend only on the fluctuation of the
levels near the Fermi energy. In order to demonstrate this, we
introduce a {\em range-$n$ approximation} in which only $n$ of the levels on either side of the
Fermi energy are left free to fluctuate, while all the others are
kept fixed at their mean position. To gauge the accuracy of this
approximation, we perform a
Kolmogorov-Smirnov test \cite{NumRecipes_KS}. In
Fig.~\ref{fig_kstest} the result of this test, namely the maximum
deviation of the range-$n$ cumulative distribution from the exact
cumulative distribution
%based on $P(|\Deltaov|^2)$
of the overlap, is
shown. Clearly, the deviation decreases to almost zero as soon as
the fluctuations of a few levels near $E_F$ are included. The convergence with $n$ appears furthermore
faster for the larger perturbation strengths in which we are interested
for the x-ray edge problem. This retrospectively
justifies the use of a full random matrix to model the statistical
properties of our chaotic systems, as the levels beyond the
Thouless energy for which such a description does not hold
%NOTE: would not apply, --> was used in PRL
do not affect the overlap distribution.

%%%%%%%%%%%%%%%%%%%%%%%%%%%%%%%
%%%%%%%%%%%%%%%%%%%%%%%%%%%%%%%%

%%%%%%%%%%%%%%%%%%%%%%%%%%%%%%%%%%%%%%%%%%%%%%%%%%%%%%%%%%%%%%%%%%%%

%%%%%%%%%%%%%%%%%%%%%%%%%%%%%%%%%%%%%%%%%%%%%%%%%%%%%%%%%%%%%%%%%%%%
%----------- begin of bound state subsection -------------------
%%%%%%%%%%%%%%%%%%%%%%%%%%%%%%%%%%%%%%%%%%%%%%%%%%%%%%%%%%%%%%%%%%%%
%\subsection{Symmetries and bound state}
%\subsection{Bound state and symmetries}
\section{Bound states and  symmetry between positive vs.~negative perturbation}
\label{sec_boundstate}
%{\it Run-away level and symmetry considerations.}

%%%%%%%%%%%%%%%%%%%%%%%%%%%%%%%
\begin{figure}
\includegraphics[width=8.5cm]{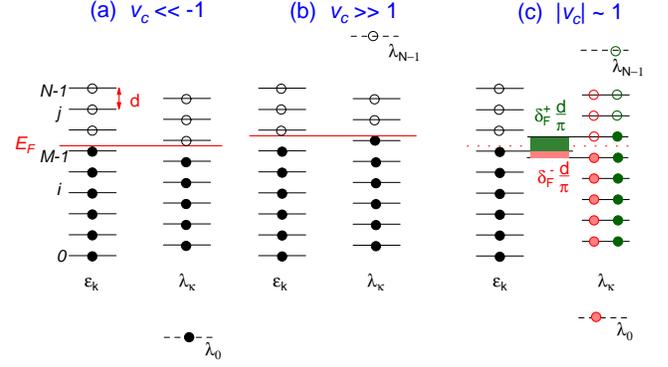}
\caption{(color online) Run-away level for strong perturbation which is
(a) positive or (b) negative. For simplicity, the bulklike situation
is shown.  In the case of intermediate perturbations (c), the
perturbed levels $\lam_{\kappa}$ are not shifted to the center
of the unperturbed level interval $\eps_k$. The phase shifts
$\delta_F^-$ (negative perturbation) and $\delta_F^+$
(positive perturbation) are related by $|\delta_F^-| +
\delta_F^+ = \pi$. In the case of strong negative perturbations, as
in the x-ray problem, $\lam_0$ forms a bound state that screens the
perturbation.
  \label{fig_symmdelf}}
\end{figure}
%%%%%%%%%%%%%%%%%%%%%%%%%%%%%%%

As soon as the perturbation strength becomes significant, i.e. for
attractive perturbations %$|\vc|/d \lesssim - 1$, the shift of the
$ \vc /d \lesssim - 1$, the shift of the lowest perturbed level,
$\lam_0 - \eps_0$, starts to become large. Analogously, for
positive perturbation strengths larger than $\vc/d \gtrsim 1$, the
same boundary effect now shifts the highest perturbed level
$\lam_{N-1}$ to very high energies, see Figs.~\ref{fig_rank1} and
\ref{fig_symmdelf}. These so-called run-away levels correspond to
the formation of a bound
state\cite{friedel_boundstate,combescotnozieres,tanabe:RMP1990,zagoskin}
which entirely screens the impurity. They can be found,
respectively, at
\begin{eqnarray}
    \lam_0 - \eps_0 &=& -\frac{N \Deltaml}{e^{d/|\vc|} - 1}
    \quad (\vc < 0)\:, \nonumber \\
    \lam_{N-1} - \eps_{N-1} &=& + \frac{N \Deltaml}{1 - e^{-d/\vc}}
    \quad (\vc > 0) \nonumber \:.
\end{eqnarray}
The fluctuations of these levels will be neglected
\cite{remark_metropolis}, since in the limit of very strong
perturbations, $|\vc|/d \gg 1$, the run-away level does not affect
the value for the overlap, Eq.~(\ref{eq:overlap}), because the
terms involving $\lam_0$ (or $\lam_{N-1}$, respectively) yield
ratios of one.  Therefore, one has only to consider the $N$
unperturbed levels $\{\eps\}$ and the $N \!-\! 1$ perturbed levels
$\lam_{\kappa}$ ``in between''. In this case of very strong
perturbations, the perturbed levels are maximally shifted: in the
center of the band, the $\lam_{\kappa}$ lie, on average, exactly in
the middle between the two neighboring unperturbed levels $\eps_{\kappa-1}$
and $\eps_{\kappa}$ (or $\eps_{\kappa}$ and $\eps_{\kappa+1}$). Therefore, a
strong negative perturbation $\vc$ (i.e.\ $\vc/d \!\ll\! -1,\;
\delta_F \!=\! -\pi/2$) and a strong positive perturbation $\vc^+$
(i.e.\ $\vc^+/d \!\gg\! 1,\; \delta_F^+ \!=\! \pi/2$) are indistinguishable, see
Fig.~\ref{fig_symmdelf}. This means in particular that the
corresponding overlap distributions $P(|\Deltaov|^2)$ and
$P(|\Deltaov^+|^2)$ are identical.

In the case of intermediate perturbation strengths $\vc/d \sim
\pm 1$, a level shift $\delta_F$ associated with the negative
perturbation corresponds to a phase shift $\delta_F^+ = \pi -
|\delta_F|$ when the level scheme is interpreted as originating
from some positive perturbation $\vc^+$ ( $\neq |\vc|$ in
general), see Fig.~\ref{fig_symmdelf}(c). We therefore expect
again the two corresponding bulk-scaled overlap distributions to
be identical, and the (symmetry) relation
\begin{equation}
  P\li(\frac{|\Deltaov|^2}{|\Deltaov_b|^2}\re)_{\vc, \delta_F}
  = P\li(\frac{|\Deltaov^+|^2}
  {|\Deltaov_b^+|^2}\re)_{\vc^+, \delta_F^+=\pi -|\delta_F|} \:
\label{eq:symmoverlap}
\end{equation}
to hold.

In Figs.~\ref{fig_symm_GOE} and \ref{fig_symm_GUE} overlap
distributions for phase shifts $0> \delta_F > -\pi$ are
shown. Starting with an almost full band ($M=98$ particles on $N=100$
levels) and a small negative phase shift, $\delta_F$ is lowered (the
effective potential strength raised) as the filling is decreased,
passing $\delta_F \approx -\pi/2$ at half filling. The corresponding
overlap distributions are shown in Fig.~\ref{fig_symm_GOE}a for the
COE situation and in Fig.~\ref{fig_symm_GUE}a for a CUE case.
% that is relevant, e.g., in the presence of a magnetic field.
In the lower panels (b) of Figs.~\ref{fig_symm_GOE} and
\ref{fig_symm_GUE} overlap distributions belonging to a negative
phase shift $\delta_F$ (and $\vc<0$) and the associated positive
phase shift $\delta_F^+ = \pi - |\delta_F|$ (and $\vc^+ >0$) are
compared and seen to (almost) coincide pairwise after scaling by
the respective bulk overlaps (the small deviations are due to the
%finite, discrete number of levels available). Note that the bulk
finite number of discrete levels available). Note that the bulk
overlaps $|\Deltaov^{+}_b|^2$ are very similar in each pair, but
rather different for different pairs, thereby causing the
stretching of the x-axis.

In the context of the x-ray edge problem (strong negative perturbation), a
situation that is equivalent to a {\it positive} perturbation %is realized
becomes of importance when the core electron is excited to the
level $\lam_M$ just above the Fermi energy $E_F$. The consequences
and more details will be discussed in
\ \onlinecite{inprep}.

%%%%%%%%%%%%%%%%%%%%%%%%%%%%%%%%%%%%%%%%%%%%%%%%%%%%%%%%%%%%%%%%%%%%

%%%%%%%%%%%%%%%%%%%%%%%%%%%%%%%%%%%%%%%%%%%%%%%%%%%%%%%%%%%%%%%%%%%%
\section{Analytic results for the overlap distribution}
%fluctuations of the AOC overlap}% and analytical results}
%\subsection{Fluctuations and Analytical Results} % for Range-1}
\label{sec_analfluctaoc}
%%%%%%%%%%%%%%%%%%%%%%%%%%%%%%%

We finish with the derivation of analytic results for the overlap
distribution.  We argued in Sec.~\ref{sec_kstest} that the
fluctuations in the ground state overlap are dominated by
fluctuations of the energy levels around the Fermi energy (see
again Fig.~\ref{fig_kstest}). We now want to deepen this
discussion and investigate to what extent an analytic
understanding of overlap probability distributions
$P(|\Deltaov|^2)$ is possible. In particular we want to understand
two characteristic qualitative features of the overlap
distribution (see for example Fig.~\ref{fig_overlapgue}): firstly
$P(|\Deltaov|^2=0)$ is finite and characteristically different for
the COE vs.~CUE situation, and secondly $P(|\Deltaov|^2=1)$ is
zero, i.e. $P(|\Deltaov|^2)$ drops to zero before the natural
bound $|\Deltaov|^2=1$ is reached.

To begin, remember that in Eq.~(\ref{eq:overlap}) for the overlap
$|\Deltaov|^2$, all energy differences are taken between empty
($j \geq M$) and occupied ($i \leq M-1$) levels.  The smallest differences
(of the order of the mean level spacing $\Deltaml$) contain the energy
levels $\eps_{M}$, $\eps_{M-1}$, and $\lam_M$. We expect therefore
that the small $\Delta$ part of the overlap distribution will be
mainly given by the fluctuation of these terms.

To gain an analytic understanding of the overlap distribution, we
therefore consider only the level closest to the Fermi energy as
fluctuating whereas all other levels are assumed at their bulklike
values (range-1 approximation).  We therefore want to evaluate
fluctuations of $(\lam_M - \eps_{M-1}) / (\eps_{M} - \eps_{M-1})$.
For simplicity, we consider the half-filled case.
Note that a similar two-level problem was studied by Vallejos {\it et al.} in
Ref.~\ \onlinecite{vallejos:PRB2002}, motivated by their observation that very small
AOC overlaps were related to avoided level-crossings at the Fermi energy.
In our present model, Eq.~(\ref{eq:lam}) for
the rank one perturbation determines the position of the perturbed
levels as solutions of the equation
\begin{equation}
f(\lam)  \df d \sum_k \frac{A_k}{\lam - \eps_k} \equiv \frac{d}{\vc}\:,
\end{equation}
where $A_k=|u_k|^2$ are the Porter-Thomas distributed wavefunction intensities,
\begin{equation}
P(A) = \left \{ \begin{array}{l}
                  e^{-A} \quad \textrm {for CUE} \\[1mm]
                  \frac{1}{\sqrt{2 \pi A}} e^{-A/2} \quad \textrm {for COE}
                \end{array} \right.\:.
\label{eq:PT}
\end{equation}
[$\la A \ra = 1$ and $\la A \ra^2 = 2 (3) $ for CUE (COE)]. We
will take only the fluctuations of the amplitudes $A_{M-1}$ and
$A_{M}$ for the wavefunctions $\phi_{M-1}$ and $\phi_M$
explicitly into account. Accordingly, we separate the first two
terms of the sum and comprise the remaining terms in a function
$\delta f$ that is assumed to be a Gaussian random variable and
contains the joint effect of all other levels and intensities,
\begin{equation}
    f(\lam) = \frac{d A_{M-1}}{\lam-\eps_{M-1}} + \frac{d A_M}{\lam-\eps_M} +
    \delta f \equiv \frac{d}{\vc}\:.
\label{eq:analterm0term1}
\end{equation}

Equation (\ref{eq:analterm0term1}) makes it possible already to
understand at a qualitative level  the finite probability for
having $|\Deltaov|^2 \!=\! 0$ which corresponds to having
$\lam_M-\eps_{M-1}=0$. However, setting $\lam \!=\! \lam_M$ in
Eq.~(\ref{eq:analterm0term1}) we see that the ratio $A_{M-1}
/ (\lam_M - \eps_{M-1}) $ has to be finite.
We conclude that having $\lam_M-\eps_{M-1}=0$ is related to having $A_{M-1} \!=\! 0$. From
Eq.~(\ref{eq:PT}) we find $P(A \!=\! 0) \!=\! 1$ [or $\infty$] in the CUE [COE] case, thereby
explaining both $P(|\Deltaov|^2 \!=\! 0) > 0$ and the peak at $|\Deltaov|^2 \!=\! 0$ in the
COE case.

To obtain a more quantitative description, we investigate $\delta f$ more closely. According to the assumptions
of the range-1 approximation, we replace the energy levels in
$\delta f$ by their bulklike values.  We furthermore neglect
boundary effects that would make the phase shift dependent on the
level number which is a good approximation for $N \to \infty$ at
half-filling.  Fixing  the energy origin in the middle of the mean
position of $\eps_{M-1}$ and $\eps_M$, we  express $\delta f$ in
the symmetric form
\begin{equation}
\delta f = \sum_{k=1}^{\infty}
           \left( \frac{A_{k+M}}{\frac{\lam}{d} - (k+\frac{1}{2})}
                + \frac{A_{-(k+1) +M}}{\frac{\lam}{d} + (k+\frac{1}{2})} \right) \:.
\end{equation}
Fluctuations in $\delta f$ are due to the fluctuations in the intensities
$A_{\pm k}$ that follow the Porter-Thomas distributions given in
Eq.~(\ref{eq:PT}). Expanding the mean value $\la \delta f \ra$ about the center
of the band, $\lam=0$, yields in lowest order
$\la \delta f \ra = - 2 \lam S_1$ %- 2 \lam^3 S_2\:,
with
\begin{equation}
S_1 = \sum_{k=1}^{\infty} \frac{1}{(k+\frac{1}{2})^2} \approx 0.9348%05
\:.
%S_2 &=& \sum_{k=1}^{\infty} \frac{1}{(n+\frac{1}{2})^4} \approx 0.234849\:.
\end{equation}
%\end{eqnarray}
%We further simplify the expression by neglecting the second term against the first one,
%and eventually arrive at
%\begin{eqnarray}
%\la \delta f \ra &=& -2 \lam S_1\\
Comprising {\it all} the randomness of the amplitudes $A_{k \neq
M\!-\!1,M}$ in one random Gaussian variable $\xsi$ with $P(\xsi) =
e^{-\xsi^2/2}/\sqrt{2\pi}$, and writing explicitly the scale
var$\, \delta f $ of the fluctuations around the mean value, we
arrive at the approximation
\begin{equation}
\delta f = \la \delta f \ra + \sqrt{\textrm{var} \,\delta f}\,\, \xsi
          = -2 \lam S_1 + \sqrt{\textrm{var}\, \delta f} \,\, \xsi
\end{equation}
with
\begin{equation}
\textrm{var} \, \delta f = \left\{ \begin{array}{l}
                        2 S_1 \quad \textrm {for CUE}\\[1mm]
                        4 S_1 \quad \textrm {for COE.}
                               \end{array}
                        \right.
\end{equation}
%Putting all together, and setting $\eps_1 - \eps_{M-1} =s$,
%the perturbed levels $\lam$ are the
%solutions of
%\begin{eqnarray}
%0&=& \frac{A_{M-1}}{\lam + \frac{s}{2}} + \frac{A_M}{\lam - \frac{s}{2}}
%- 2 \lam S_M + \textrm{var}\, \delta f \,\, \xsi - \frac{1}{\vc d N}
%\label{eq:rank1anal_M} \\
%&\df&\Lambda(A_{M-1},A_M,\xsi)|_{s,\vc} \nonumber \:.
%\end{eqnarray}
%Eventually,
Setting $\eps_M - \eps_{M-1} =s$, the probability distribution
$P(\lam|s)$ is obtained by integrating over the distributions of
$A_{M-1},A_M$ and
$\xsi$. These
integrals can only partially be solved analytically and,
%\begin{equation}
%P(\lam|s) = \int dA_{M-1} P(A_{M-1}) dA_M P(A_M) d\xsi P(\xsi)
%\delta (\lam - \Lambda(A_{M-1},A_M,\xsi))\:.
%\end{equation}
%Since for each $A_{M-1},A_M$, and $\lam$ there exists a solution $\xsi=\xsi_{M-1}$ for
%Eq.~(\ref{eq:rank1anal_M}), we can re-write the last equation
%\begin{equation}
%P(\lam|s) = \int dA_{M-1} dA_M P(A_{M-1}) P(A_M) P(\xsi_{M-1})
% \li|\frac{\p f}{\p \lam}\re| \li|\frac{\p f}{\p \xsi}\re|^{-1}
%\end{equation}
%where
%\begin{equation}
%\xsi_{M-1} = \frac{1}{\sqrt{\textrm {var} \, \delta f}}
%\li( \frac{1}{\vc d N} + 2 S_M \lam-\frac{A_{M-1}}{\lam+\frac{s}{2}}-
%\frac{A_M}{\lam-\frac{s}{2}} \re)\:.
%\end{equation}
leaving out the further details of the calculation, the result for the CUE case ($\mu=\alp_M-\alp_{M-1},
\alp_{0} =A_{M-1}/(\lam+s/2), \alp_{1} = - A_M/(\lam-s/2)$) is
\begin{widetext}
% \begin{eqnarray}
% P(\lam|s) &=& \frac{1}{2s\sqrt{\pi S_1}} \int_{-\infty}^{\infty} d\mu
%                         \li\{
%                         \exp{\li[ \mu \lam - %wrong was "+"
%                         |\mu|\frac{s}{2}\re]}\nonumber \re.\\
% &&\times \exp{\li[- \frac{1}{4S_1}\li(2S_1\lam + \frac{1}{\vc} +\mu\re)^2\re]} \label{eq:PanalCUE}\\
% &&\times \li. \li[ 2 S_1 \li( \frac{s^2}{4} - \lam^2 \re) +
%  \mu \lam + |\mu| \frac{s}{2} + 1\re]
%                         \re\} \nonumber \:,
% \end{eqnarray}
\begin{equation}\label{eq:PanalCUE}
P(\lam|s) = \frac{1}{2s\sqrt{\pi S_1}} \int_{-\infty}^{\infty} \!\!d\mu
                        \li\{
                        \exp{\li[ \mu \lam - %wrong was "+"
                        |\mu|\frac{s}{2}\re]}
\exp{\Big[-\frac{1}{4S_1}\Big(2S_1\lam + \frac{1}{\vc} +\mu\Big)^2\Big]}
 \Big[ 2 S_1 \Big( \frac{s^2}{4} - \lam^2 \Big) +
 \mu \lam + |\mu| \frac{s}{2} + 1\Big]
                        \re\}  \:,
\end{equation}
and for the COE case
% \begin{eqnarray}
% P(\lam|s) &=&
% % do some cosmetics here \frac{1}{(2\pi)^{3/2}} \frac{1}{\sqrt{4 S_1}
% %\sqrt{\frac{s^2}{4} - x^2}}\nonumber \\
% \frac{1}{\sqrt{ (2\pi)^{3} \, 4 S_1 \, \li( \frac{s^2}{4} - \lam^2 \re)} } \nonumber \\
%   &\times& \int_{-\infty}^{\infty} d\mu \li[
%         \exp{\li[-\frac{1}{8S_1} \li(2 S_1 \lam +\frac{1}{\vc} +
%             \mu\re)^2 +
%             \mu \frac{\lam}{2}\re]} \re. \nonumber \\
%   &\times& \li\{ \li[2 S_1 \li(\frac{s^2}{4} -\lam^2\re)+\mu \lam \re]
%         K_0\li(\li|\frac{s\mu}{4}\re|\re) \re.\nonumber \\
%               &&  \li. \li.  + 2 \li|\frac{s\mu}{4}\re|
%         K_1\li(\li|\frac{s\mu}{4}\re|\re)\re\} \re]\:,  \label{eq:PanalCOE}
% \end{eqnarray}
\begin{eqnarray}
P(\lam|s) =
\frac{1}{\sqrt{ (2\pi)^{3} \, 4 S_1 \, \li( \frac{s^2}{4} - \lam^2 \re)} }
& \, & \int_{-\infty}^{\infty} \!\! d\mu \li[
        \exp{\Big[-\frac{1}{8S_1} \Big(2 S_1 \lam +\frac{1}{\vc} +
            \mu\Big)^2 +
            \mu \frac{\lam}{2}\Big]} \re.  \label{eq:PanalCOE} \\
  &\times& \li. \li\{ \Big[2 S_1 \Big(\frac{s^2}{4} -\lam^2\Big)+\mu \lam \Big]
        K_0\Big(\Big|\frac{s\mu}{4}\Big|\Big)
               + 2 \Big|\frac{s\mu}{4}\Big|
        K_1\Big(\Big|\frac{s\mu}{4}\Big|\Big)\re\} \re]\:, \nonumber
\end{eqnarray}
with $K_0$, $K_1$ the modified Bessel functions.
%CHECK: Improve the result using the error function!!!ENDCHECK
%In both cases, this last integral over $\mu$ has to be performed numerically,
%whereas the other integrals could be solved analytically in terms of Bessel
%functions $K_0$ and $K_1$.

The remaining integral over $\mu$ can be performed in the CUE case
using the (inverse) error function. Splitting the
$\mu$-integration into integration from 0 to $\infty$ and from
$-\infty$ to 0, yielding results $P_+(\lam|s)$ and $P_-(\lam|s)$,
%of the $\mu$-integration from 0 to $\infty$ and $-\infty$ to 0,
%respectively, read with
and using the substitutions $A=2
\,S_1\, \lam \;+\;1/\vc$, $B=2\,S_1\,(s^2/4 -\lam^2)$, $a=1/(4
S_1)$, and $b_{\pm} = [s \pm 1/(S_1 \vc)]/2$ we find
% \begin{eqnarray}
% P_{\pm}(\lam|s) & = & \frac{\exp(-A^2 a)}{2\,a\,C}
%                     \li\{ \pm \li(\lam \pm \frac{s}{2} \re) \re. \nonumber \\
% &&                  \li. +\sqrt{\frac{\pi}{a}} \; \exp \li( \frac{b_{\pm}^2}{4a} \re) \;
%                       {\rm erfc} \li( \frac{b_{\pm}}{2\, \sqrt{a}} \re) \re. \nonumber \\
% &&                  \li. \times \li[ B\, a \mp \frac{b_{\pm}}{2} \li(\lam \pm \frac{s}{2} \re) \re] \re\} \:,
% \end{eqnarray}
\begin{equation}
P_{\pm}(\lam|s) =  \frac{\exp(-A^2 a)}{2\,a\,C}
                    \li\{ \pm \li(\lam \pm \frac{s}{2} \re)
  +\sqrt{\frac{\pi}{a}} \; \exp \li( \frac{b_{\pm}^2}{4a} \re) \;
                      {\rm erfc} \li( \frac{b_{\pm}}{2\, \sqrt{a}} \re)
    \Big[ B\, a \mp \frac{b_{\pm}}{2} \li(\lam \pm \frac{s}{2} \re) \Big]
    \re\} \:,
\end{equation}
and their sum yields $P(\lam|s) = P_+(\lam|s) + P_-(\lam|s)$.

This expression simplifies considerably
% A simplification is\ furthermore possible
in the limit of very strong perturbations, $|\vc| \gg 1$. Then, the
result becomes independent of $\vc$ (in particular independent of
the sign -- very strong attractive and repulsive interactions are
equivalent), and takes the form
% \begin{eqnarray}
% %P_{\infty} (\lam|s)
% %& = &\frac{e^{-\lam^2 S_1}}{s \sqrt{\pi}}
% %\li[ s \sqrt{S_1}  \re.  \label{eq:probr1infty} \\
% %& & \li. + \sqrt{\pi} e^{s^2 S_1/ 4} \li(1 - 2 \lam^2 S_1\re) \,
% % {\textrm {erfc} \li(\frac{s \sqrt{S_1}}{2} \re)} \re] \:, \nonumber
% P_{\infty} (\lam|s)
% & = &\frac{\exp(-\lam^2 S_1)}{s \sqrt{\pi}}
% \li[ s \sqrt{S_1}  \re.  \label{eq:probr1infty} \\
% & & \li. + \sqrt{\pi} \exp\li(\frac{s^2 S_1}{4}\re)
%  {\rm erfc} \li( \frac{s \sqrt{S_1}}{2} \re)  \re. \nonumber \\
% && \li. \times \li(1 - 2 \lam^2 S_1\re) \re]  \:. \nonumber
% \end{eqnarray}
\begin{equation} \label{eq:probr1infty}
P_{\infty} (\lam|s)
= \frac{\exp(-\lam^2 S_1)}{s \sqrt{\pi}}
\li[ s \sqrt{S_1}  + \sqrt{\pi} \li(1 - 2 \lam^2 S_1\re)
 \exp\li(\frac{s^2 S_1}{4}\re)
 {\rm erfc} \li( \frac{s \sqrt{S_1}}{2} \re)
  \re]  \:.
\end{equation}
\end{widetext}
Clearly, this probability distribution is symmetric with respect
to $\lam=0$ as it should be.
%%%%\end{widetext}

The contribution to the overlap, Eq.~(\ref{eq:overlap}), that contains the
energy levels closest to the Fermi energy is represented by
$(\lam_M-\eps_{M-1})/(\eps_M-\eps_{M-1}) \equiv (\lam + s/2)/s $ in our
present notation.
To obtain the probability distribution $P^{F1} (\lam)$ of this quantity,
%To arrive at the final result of our range-1 approximation
%Finally, %(numerical) integration over the
we have to take into account the distribution of nearest
neighbor spacings $s=\eps_M-\eps_{M-1}$ according to the
Wigner surmise,
\begin{equation}
P(s) = \left\{ \begin{array}{l}
%                \frac{32}{\pi^2}\; s^2 \; e^{-\frac{4}{\pi} s^2}
                \frac{32}{\pi^2}\; s^2 \; \exp \li( -\frac{4}{\pi} s^2 \re)
                \quad \textrm {for CUE}\\[1.5mm]
%                \frac{\pi}{2}\; s \; e^{-\frac{\pi}{4} s^2}
                \frac{\pi}{2}\; s \; \exp \li( -\frac{\pi}{4} s^2 \re) \quad \textrm {for COE}\:.
                \end{array} \right.
\end{equation}
The final result then reads
\begin{equation}
P^{F1}(\lam) = %\int_{\lam}^{\infty} ds P(s) P(\lam|s)\:.
                \int_{0}^{\infty} ds \,\, s \, P(\lam|s) \, P(s)\:,
\end{equation}
This integrated result is however essentially indistinguishable
from $P(\lam|s=1)$.

\begin{figure}
\includegraphics[width=8.5cm]{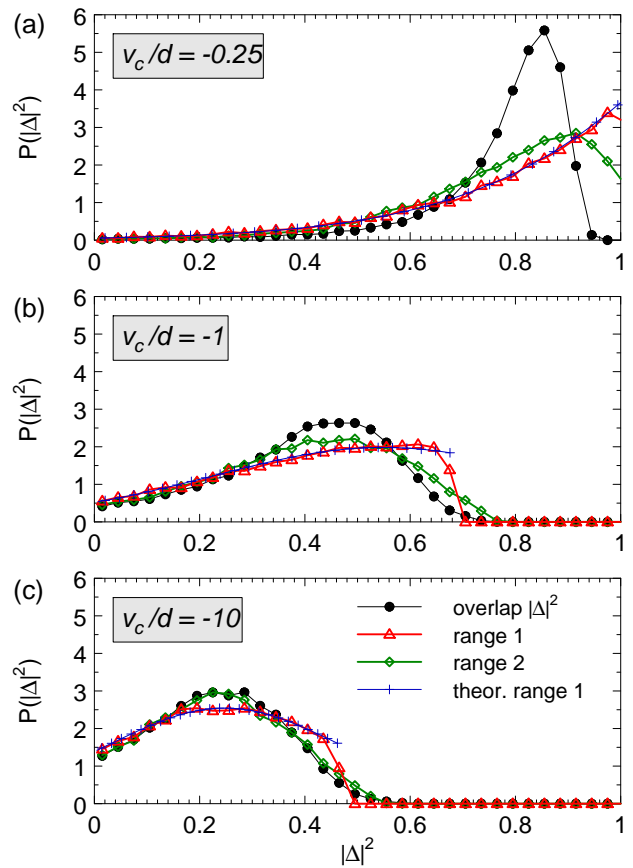}
\caption{(color online) Probability distributions $P(|\Deltaov|^2)$ for
the ground state overlap ($N=100, M=50$, CUE) for increasing potential
strength: (a) weak ($\vc/d = -0.25$), (b) intermediate
($\vc/d = -1$), and (c) strong ($\vc/d = -10$) perturbation. Shown are
the exact distribution (dots) as well as the range-1 (triangles) and range-2
(diamonds) approximations. Clearly, the fluctuations come from the
levels closest to the Fermi energy, and the range-1 approximation
provides a good estimate already for intermediate perturbation
strengths as in (b). The crosses denote the analytic result for
the range-1 approximation, showing excellent agreement with the
numerical results.
  \label{fig_overlfluct_anal_GUE}}
\end{figure}

The probability distribution of the range-1 approximation,
$P^{r1}(|\Delta^{r1}|^2)$, is
straightforwardly obtained from $P^{F1}(\lam)$
by scaling with the bulklike quantity $|\Deltaov_b|^2 / (1 - \delta_F/\pi)$
[cf.~Eq.~(\ref{eq:range1bulknorm}) below] and shifting the argument by 1/2.
%These analytic results are presented in Fig.~\ref{fig_overlfluct_anal_GUE}
The resulting curves are presented in
Fig.~\ref{fig_overlfluct_anal_GUE} for weak, intermediate, and strong
perturbations.  This analytic
range-1 result is compared to the full overlap distribution
$P(|\Deltaov|^2)$ as well as the exact range-1 and
range-2 approximations\cite{remark_range2} (CUE case,
see Ref.~\ \onlinecite{xrayprl} for the corresponding COE figure).
The agreement between the analytic approximation and the exact
range-1 results is excellent. The analytic range-1 result
provides in particular a very good estimate for the full overlap
distribution at small overlaps, independent of the perturbation
strength. Moreover, in the limit of strong perturbations in which we
are especially interested for the x-ray edge problem,
it provides a reasonable estimate of the full probability
distribution $P(|\Deltaov|^2)$. The increase in the degree of
symmetry of the range-1 result (with respect to  the maximum of
the curve) as the perturbation strength is increased is clearly
visible and confirmed by Eq.~(\ref{eq:probr1infty}).

%%%%%%%%%%%%%%%%%%%%%%%%%%%%%%%%

We would like to mention that Vallejos and coworkers in
Ref.~\ \onlinecite{vallejos:PRB2002}, and, based on their work,
Gefen and coworkers in Ref.~\ \onlinecite{gefen:PRBR2002}, found
similar results, in particular that the distribution of
small-overlaps is well represented by a two-level analytic
model. However, there are few differences. For example, the
difference between the (Gaussian) orthogonal and unitary case was
less pronounced in their models, especially the increase of the
probability $P(|\Deltaov|^2)$ towards $|\Deltaov|^2 \to 0$,
characteristic of our model, was not seen. Further, simple
analytic expressions similar to the expressions
Eqs.~(\ref{eq:PanalCUE})-(\ref{eq:PanalCOE}) could only be found
for the orthogonal situation in
Ref.~\ \onlinecite{vallejos:PRB2002}.
% Vallejos {\it et al.} found the
%Gaussian orthogonal ensemble to give simpler analytic results
%than the unitary case for their model. huiohuio

Eventually, we briefly remark on having $P(|\Deltaov|^2=1)=0$ for finite
perturbation strength. Within a range-1-like approximation scheme,
Eq.~(\ref{eq:overlap}) can be written as
\begin{equation}
|\Deltaov^{r1}|^2 = \frac{(\lam_M - \eps_{M-1})}{(\lam_M - \lam_{M-1})}\:
                    \frac{\Deltaml}{\Deltaml \li(1 - \frac{\delta_F}{\pi} \re)}
                    |\Deltaov_b|^2\:,
\label{eq:range1bulknorm}
\end{equation}
where we have normalized all differences by their bulklike values.
Only the first term is fluctuating, between 0 and 1, whereas the
last two terms are numbers that are strictly smaller than 1 and, furthermore,
get smaller as the perturbation gets stronger. Accordingly, the
upper boundary 1 of the first term is shifted to a value
$|\Deltaov_{\rm max}^{r1}|^2 <1 $ that decreases with increasing
perturbation strength as is clearly visible in
Fig.~\ref{fig_overlfluct_anal_GUE}. The value $|\Deltaov_{\rm
max}^{r1}|^2 $ provides a good estimate for the upper boundary of
the overlap $|\Deltaov|^2$, although there are of course
corrections from higher order approximations that smear the
(total) overlap distribution $P(|\Deltaov|^2)$ about
$|\Deltaov_{\rm max}^{r1}|^2 $, see for example
Fig.~\ref{fig_overlfluct_anal_GUE}(b).

%%%%%%%%%%%%%%%%%%%%%%%%%%%%%%%%%%%%%%%%%%%%%%%%%%%%%%%%%%%%%%%%%%%%
%% Summary
%%%%%%%%%%%%%%%%%%%%%%%%%%%%%%%%%%%%%%%%%%%%%%%%%%%%%%%%%%%%%%%%%%%%
%%%%%%%%%%%%%%%%%%%%%%%%%%%%%%%%%%%%%%%%%%%%%%%%%%%%%%%%%%%%%%%%%%%%

\section{Summary and Conclusions}

\label{sec_summary}
%%%%%%%%%%%%%%%%%%%%%%%%%%%%%%%%%%%%%%%%%%%%%%%%%%%%%%%%%%%%%%%%%%%%

In this paper we have studied the Anderson orthogonality catastrophe
in generic mesoscopic systems with a finite number of electrons in
discrete energy levels.  Motivated by the study of mesoscopic
photoabsorption spectra that will be the subject of the second
paper \ \onlinecite{inprep} in this series, we have used the model
of a rank one perturbation describing the perturbation left behind
after the photoexcitation of a core electron into the conduction
band. However, the basic underlying physics is much more general,
and the model can be applied to any situation where a sudden,
localized (rank one) perturbation acts on a finite number of
chaotic electrons.

As a result of such a sudden perturbation, the overlap between the
unperturbed and perturbed many-body ground states vanishes in the
thermodynamic limit (AOC).  In mesoscopic systems, even with few
thousand electrons, this effect will be incomplete. Fluctuations
cause a broad distribution of overlaps that is, however, bounded
from above by a value smaller than 1 (for not too small
perturbations) whereas there is a non-zero probability for having
zero-overlap. The distribution of small overlaps is distinctly
different for systems with or without time reversal symmetry (COE
vs.~CUE) and traces back to the difference in the Porter-Thomas
distribution of their single-particle wavefunction intensities.
The fluctuations originate from the levels around the Fermi energy
(within a range smaller than the Thouless energy for systems we
are interested in), and we derived an analytic expression for
the lowest order (range 1) result that reproduces the distribution
of small overlaps and is a good approximation for the whole
distribution in case of strong perturbations.

\begin{acknowledgements}
We thank K.~Matveev for several helpful discussions and
I.~Aleiner, Y.~Gefen, I.~Lerner, E.~Mucciolo, and I.~Smolyarenko
for useful conversations. M.~H.~acknowledges the support of the
Alexander von Humboldt Foundation.  This work was supported in part
by the NSF (DMR-0103003).
\end{acknowledgements}

%%%%%%%%%%%%%%%%%%%%%%%%%%%%%%%%%%%%%%%%%%%%%%%%%%%%%%%%%%%%%%%%%%%%
%%%%%%%%%%%%%%%%%%%%%%%%%%%%%%%%%%%%%%%%%%%%%%%%%%%%%%%%%%%%%%%%%%%%


\begin{thebibliography}{99}

\bibitem[II]{inprep}
M.~Hentschel, D.~Ullmo, and H.~U.~Baranger, in preparation.

\bibitem{bulk_experiments_metal}  %Experiments for bulk Xray.n
For a comprehensive experimental review on the metallic case, see
P.~H.~Citrin, G.~K.~Wertheim, and M.~Schl\"uter, Phys.~ Rev.~B
{\bf 20}, 3067 (1979) and references therein.

\bibitem{bulk_experiments_semi}
J.~S.~Lee, Y.~Iwasa, and N.~Miura,
Semicond.~Sci.~Technol.~{\bf 2}, 675 (1987);
M.~S.~Skolnick {\it et al.},
Phys.~Rev.~Lett.~{\bf 58}, 2130 (1987).

\bibitem{nozieres} B.~Roulet, J.~Gavoret, and P.~Nozi\`{e}res,
Phys.~Rev. {\bf 178}, 1072 (1969);
P.~Nozi\`{e}res,  J.~Gavoret, and B.~Roulet,
Phys.~Rev. {\bf 178}, 1084 (1969);
P.~Nozi\`{e}res and C.~T.~De Dominicis,
Phys.~Rev. {\bf 178}, 1097 (1969).

\bibitem{mahan:book} G.~D.~Mahan, \emph{Many-Particle Physics}, 2nd edition,
pp. 732 (Kluwer Academic/Plenum Publishers, New York, 1993). %3rd edition: 2000.

\bibitem{schotteschotte}
K.~D.~Schotte and U.~Schotte, Phys.~Rev. {\bf 182}, 479 (1969).

\bibitem{tanabe:seriesofpapers} %Series of papers by Tanabe and Othaka:
K.~Ohtaka and Y.~Tanabe, Phys. Rev. B {\bf 28}, 6833 (1983);
%\{I: Golden rule approach\}
Y.~Tanabe and K.~Ohtaka, Phys. Rev. B {\bf 29}, 1653 (1984);
%\{II: Effect of a bound state\}
K.~Ohtaka and Y.~Tanabe, Phys. Rev. B {\bf 30}, 4235 (1984);
%\{III: Temperature dependence\}
Phys. Rev. B {\bf 34}, 3717 (1986);
%\{IV: Numerical analysis\}
Phys. Rev. B {\bf 39}, 3054 (1989).
%\{V: Thermal broadening, Comparison with Exp. in quantum wells\}

\bibitem{tanabe:RMP1990}  Theoretical work on the metallic x-ray edge problem
%References \onlinecite{nozieres,mahan:book,tanabe:seriesofpapers}
is reviewed in K.~Ohtaka and Y.~Tanabe, Rev.~Mod.~Phys.~{\bf 62},
929 (1990).

\bibitem{xrayprl} M.~Hentschel, D.~Ullmo, and H.~U.~Baranger,
Phys. Rev. Lett. {\bf 93}, 176807 (2004).

\bibitem{sohn:MesoBook1997}
L.~L.~Sohn, G.~Sch\"on, and L.~P.~Kouwenhowen,
\emph{Mesoscopic Electron Transport} (Kluwer, Dordrecht, 1997).

\bibitem{beenakker:RMP1997}
C. W. J. Beenakker, Rev. Mod. Phys. \textbf{69}, 731 (1997).

\bibitem{alhassid:RMP2000}
Y. Alhassid, Rev. Mod. Phys. \textbf{72}, 895 (2000).

\bibitem{anderson:PRL1967}
P.~W.~Anderson, Phys.~Rev.~Lett.~{\bf 18}, 1049 (1967).

\bibitem{moessbauer} R.~L.~M\"ossbauer and D.~H.~Sharp,
 Rev.~Mod.~Phys. {\bf 36}, 410 (1964).

\bibitem{glazmanetal} %Glazman Double Dot
K.~A.~Matveev, L.~I.~Glazman, and H.~U.~Baranger,
Phys.~Rev.~B {\bf 54}, 5637 (1996).

\bibitem{levitov} D.~A.~Abanin and L.~S.~Levitov,
Phys.~Rev.~Lett. {\bf 93}, 126802 (2004).

\bibitem{matveev:larkin}
K.~A.~Matveev and A.~I.~Larkin, Phys.~Rev.~B {\bf 46}, 15337 (1992).

\bibitem{haug}
I.~Hapke-Wurst, U.~Zeitler, H.~Frahm, A.~G.~M.~Jansen, R.~J.~Haug, and K.~Pierz, % {\it et al.},
Phys.~Rev.~B. {\bf 62}, 12621
(2000) and Refs. [5]-[7] therein.

\bibitem{calleja}
J.~M.~Calleja, A.~R.~Go\~{n}i, B.~S.~Dennis, J.~S.~Weiner, A.~Pinczuk, S.~Schmitt-Rink,
L.~N.~Pfeiffer, K.~W.~West, J.~F.~M\"uller, and A.~E.~Ruckenstein,
%{\it et al.},
Sol.~St.~Comm. {\bf 79}, 911 (1991).

\bibitem{oreg:PRB1996}
Y.~Oreg and A.~M.~Finkelstein,
Phys.~Rev.~B {\bf 53}, 10928 (1996).

\bibitem{kroha:PRB1992}
Y.~Chen and J.~Kroha, Phys.~Rev.~B {\bf 46}, 1332 (1992).
%\bibitem{doniach_from_kroha} S.~Doniach and M.~\v{S}unji\'{c}, J.~Phys.~C {\bf 3},
%285 (1970).

\bibitem{matveev:PRL1998}
I.~L.~Aleiner and K.~A.~Matveev, Phys.~Rev.~Lett.~{\bf 80}, 814 (1998).

\bibitem{vallejos:PRB2002}
R.~O.~Vallejos, C.~H.~Lewenkopf, and Y.~Gefen,
Phys.~Rev.~B {\bf 65}, 085309 (2002).

\bibitem{gefen:PRBR2002}
Y.~Gefen, R.~Berkovits, I.~V.~Lerner, and B.~L.~Altshuler,
Phys.~Rev.~B {\bf 65}, 081106(R) (2002).

\bibitem{Oriol:Houches}
O.~Bohigas, in \emph{Chaos and Quantum Physics} (Les Houches Session LII, 1989),
edited by M.-J.~Giannoni, A.~Voros, and J.~Zinn-Justin (North Holland, 1991),
pp.~87.

\bibitem{mehta}
M.~L.~Mehta, \emph{Random Matrices,} 2nd edition (Academic Press, San Diego, 1991).

\bibitem{smolyarenko:PRL2002}
I.~E.~Smolyarenko, F.~M.~Marchetti, and B.~D.~Simons,
Phys.~Rev.~Lett.~{\bf 88}, 256808 (2002).

\bibitem{berry1985}
M. V. Berry, Proc. R. Soc. London \textbf{400}, 229 (1985).

\bibitem{altshuler1986}
B. L. Altshuler and B. I. Shklovskii, Zh. Eksp. Teor. Fiz. \textbf{91},
220 (1986) [Sov. Phys. JETP \textbf{64}, 127 (1986)].

\bibitem{NumRecipes_KS}
W.~H.~Press, S.~A.~Teukolsky, W.~T.~Vetterling, B.~P.~Flannery,
\emph{Numerical Recipes in C}
(Cambridge University Press, 1992), p.~623.

\bibitem{friedel_boundstate}
J.~Friedel, Phil. Mag. {\bf 43}, 153 (1952).

\bibitem{combescotnozieres}
M.~Combescot and P.~Nozi\`{e}res, J.~Phys.~{\bf 32}, 913 (1971). % Journal de Physique

\bibitem{zagoskin}
A.~M.~Zagoskin and I.~Aflleck, J.~Phys.~A {\bf 30}, 5743 (1997).

\bibitem{remark_metropolis} When the Metropolis algorithm is interrupted after
a sufficient number of steps, the circle is ``opened'', and the runaway level
is introduced by hand, neglecting possible fluctuations.

\bibitem{remark_range2} The range-1 and range-2 approximation in
Fig.~\ref{fig_overlfluct_anal_GUE} are based on
Eq.~(\ref{eq:range1bulknorm}) and its generalization to include
differences up to two mean level spacings $d$ on average for the
range-2 result.





\end{thebibliography}
\end{document}